\documentstyle[aps,prl,graphicx]{revtex}

\input psfig.sty

\newcommand{\scrs}[1]{\mbox{\scriptsize{#1}}}

\begin{document}
\twocolumn
\wideabs{

\bibliographystyle{prsty}  

\title{Loading a vapor cell magneto-optic trap using light-induced atom desorption}

\author{B.P. Anderson\cite{BPApresentaddress} and M.A. Kasevich}
\address{Physics Department, Yale University, New Haven,  CT 06520-8120}
\date{July 14, 2000}

\maketitle

\begin{abstract}

Low intensity white light was used to increase the loading rate of
$^{87}$Rb atoms into a vapor cell magneto-optic trap by inducing
non-thermal desorption of Rb atoms from the stainless steel walls
of the vapor cell.  An increased Rb partial pressure reached a
new equilibrium value in less than 10 seconds after switching on
the broadband light source. After the source was turned off, the
partial pressure returned to its previous value in $1/e$ times as
short as 10 seconds.\\

PACS number(s): 32.80.Pj, 42.50.Vk, 68.45.Da\\
\end{abstract}

}
\par

\section{Introduction}

The evaporative cooling techniques used to achieve Bose-Einstein
condensation in atomic gases \cite{and95,dav95,bra95,fri98} rely on loading large
numbers of atoms into magnetic traps with long trap lifetimes. The
approach originally taken by Anderson {\it et al.}~\cite{and95}
was to load Rb atoms into a vapor cell \cite{mon90} magneto-optic
trap (MOT) \cite{raa87JOSAB89} and to subsequently transfer atoms
into a magnetic trap located in the same cell. Large numbers of
atoms and long lifetimes were achieved by optimizing the Rb
partial
pressure and by working with long MOT loading times.

We have found a simple way to improve such a setup by modulating
the vapor pressure such that it is high for initial trap loading
and then low again in order to achieve long lifetimes in a
magnetic trap.  The technique requires the use of a white light
source (WLS) with radiation incident upon the inner walls of the
vapor cell.  When such a light source is turned on, Rb atoms that
coat the inner walls of the stainless steel vacuum chamber are
quickly desorbed and the Rb vapor pressure suddenly increases. The
vapor pressure soon returns to equilibrium after the WLS is turned
off. This enables loading large numbers of atoms into the MOT in a
relatively short amount of time, while preserving the low
pressures required for long magnetic trap lifetimes.  The WLS
method that we describe here is a possible alternative to the
double-chamber techniques \cite{mya96} and Zeeman slowing
techniques \cite{dav95,bra95} currently used to capture atoms
before evaporatively cooling in a magnetic trap.  In our
experiments, the
WLS method is used in the manner described here for the achievement of BEC in a vapor cell,
where the WLS frees us from environmentally induced variations in
vapor pressure; for example, regardless of chamber temperature, we
can load large numbers of atoms into our MOT and achieve BEC \cite{and99}.

Light-induced atom desorption (LIAD) has previously been used to
obtain optically thick Na and Rb vapors in cells made of glass,
pyrex, and sapphire \cite{bon90,meu94xu87mar94}.  In most of these
experiments, the inner walls of the vapor cells were coated with
paraffin or silane in order to enhance the LIAD efficiency by
reducing the alkali atom adsorption energy \cite{adsorbgeneral}.
In our work, an optically thick vapor was not required.  Since we
did not need desorption rates characteristic of coated cells, we
could desorb atoms directly from stainless steel.

\section{Background}

We first review the basic mechanisms involved in the operation of
a vapor cell MOT, lucidly described in Ref.~\cite{mon90}, in order
to understand the gains available with the LIAD method.  In a
vapor cell MOT, atoms are loaded into the MOT at a rate $R$. This
rate depends upon the size and intensity of the laser cooling and
trapping beams and the Rb partial pressure. Atoms with velocities
below a critical velocity will be captured by the trap. Atoms are
also lost from the trap due to collisions, limiting the number
that can be loaded into the MOT. The rate equation for the number,
$N$, of trapped atoms is given by
\begin{equation}
 \frac{dN}{dt} = R - N\left(\frac{1}{\tau_{\scrs{b}}}+\frac{1}{\tau_{\scrs{Rb}}}\right)-
 \beta \int n^2 dV,
\end{equation}
where $1/\tau_{\scrs{b}}$ is the trap loss rate due to collisions
with background gas atoms and $1/\tau_{\scrs{Rb}}$ is the loss
rate determined by collisions with untrapped Rb atoms.  The trap
density, $n$ in the volume integral, contributes to density-dependent
losses within the trap with a loss coefficient of
$\beta$. The loss rate $1/\tau_{\scrs{b}}$ is proportional to the
pressure of the background gas, and like $R$, $1/\tau_{\scrs{Rb}}$
is proportional to the Rb partial pressure. In the absence of
density-dependent collisional losses ($\beta=0$) \cite{wal94}, and
with a Rb partial pressure that is much higher than the background
pressure ($1/\tau_{\scrs{Rb}} \gg 1/\tau_{\scrs{b}}$), the rate
equation becomes
\begin{equation}
 \frac{dN}{dt} = R - \frac{N}{\tau_{\scrs{Rb}}}.
 \label{eq:liadrateeq}
\end{equation}
The limiting number, $N_{\scrs{lim}}$, that can be loaded into the
MOT is obtained when the increase in number due to loading
balances the loss due to collisions. At this point,
$N=N_{\scrs{lim}}$ and $dN/dt=0$, yielding
\begin{equation}
 N_{\scrs{lim}}=R\tau_{\scrs{Rb}},
\end{equation}
independent of the Rb partial pressure \cite{mon90}.

Frequently, the background-gas collisions can not be neglected,
and the total number reached will be less than $N_{\scrs{lim}}$.
The maximum number $N_{\scrs{max}}$ that can be captured for a
given Rb partial pressure will then be given by
\begin{equation}
 N_{\scrs{max}} = N_{\scrs{lim}}\frac{\tau_{\scrs{MOT}}}{\tau_{\scrs{Rb}}}=R\tau_{\scrs{MOT}},
 \label{eq:liadN0}
\end{equation}
where
\begin{equation}
 \frac{1}{\tau_{\scrs{MOT}}}=\frac{1}{\tau_{\scrs{b}}}+\frac{1}{\tau_{\scrs{Rb}}}.
\end{equation}
If the trap starts filling at time $t=0$, the number of atoms in
the MOT at any point in time is given by
\begin{equation}
 N(t)=N_{\scrs{max}}\left[1-\exp{\left(-\frac{t}{\tau_{\scrs{MOT}}}\right)} \right].
 \label{eq:liadloading}
\end{equation}
Because of the appearance of $\tau_{\scrs{MOT}}$ as the time
constant in the exponential, we define $\tau_{\scrs{MOT}}$ as the
''MOT loading time.''

The lifetime of a \emph{magnetic trap} in the same chamber also
depends upon the collision rate of trapped atoms with background
atoms. Thus the magnetic trap lifetime $\tau$ is proportional to
$\tau_{\scrs{MOT}}$.  We express this proportionality as
$\tau=\tau_{\scrs{MOT}}/\alpha$, where for our traps, $\alpha \sim
4$.  For evaporative cooling experiments, where large numbers of
atoms and long magnetic trap lifetimes are both necessary, the
product of total number~$N_{\scrs{max}}$ and magnetic trap
lifetime is the critical parameter to maximize \cite{and94b}.
Because of the relationship between $\tau$ and
$\tau_{\scrs{MOT}}$, we can alternatively view this requirement as
maximizing the product of $N_{\scrs{max}}$ and
$\tau_{\scrs{MOT}}$. We must therefore find the optimum Rb
partial pressure for a given background pressure. Multiplying
Eq.~\ref{eq:liadN0} by $\tau_{\scrs{MOT}}$ leads to
maximization of $\tau_{\scrs{MOT}}^2/\tau_{\scrs{b}}$ ($N_{\scrs{lim}}$
is independent of vapor pressure).  Under
optimal conditions, with constant Rb partial pressure,
$N_{\scrs{max}}\tau_{\scrs{MOT}}$ is maximized for
$\tau_{\scrs{b}}=\tau_{\scrs{Rb}}=\tau_{\scrs{MOT}}/2$ and hence
$N_{\scrs{max}}=N_{\scrs{lim}}/2$ and
$N_{\scrs{max}}\tau_{\scrs{MOT}}=N_{\scrs{lim}}\tau_{\scrs{b}}/4$.

However, we can further improve the number-lifetime product (which
from now on we will generally designate as $N\tau$) by permitting
a modulation of the Rb vapor pressure. If the Rb partial pressure
is \emph{temporarily} increased until the trap contains
the maximum possible number of atoms %
($N = N_{\scrs{max}} = N_{\scrs{lim}}$), at which point the Rb
vapor is suddenly reduced to a negligible level
($\tau_{\scrs{MOT}}\sim \tau_{\scrs{b}}$), an increase of a factor
of 4 in $N\tau$ will be realized. Furthermore, the time needed to
load the MOT is significantly shortened when
$\tau_{\mbox{\scriptsize{Rb}}}
\ll \tau_{\scrs{b}}$ during the loading interval, increasing the
repetition rate of the experiment.

The goal of our experiment was to realize gains in $N\tau$ by
modulating the Rb vapor pressure in the described manner with the
white light source, thus improving conditions for evaporative
cooling and obtaining BEC.

\section{Experimental setup and measurement techniques}

\begin{figure}     
\begin{center}
\psfig{figure=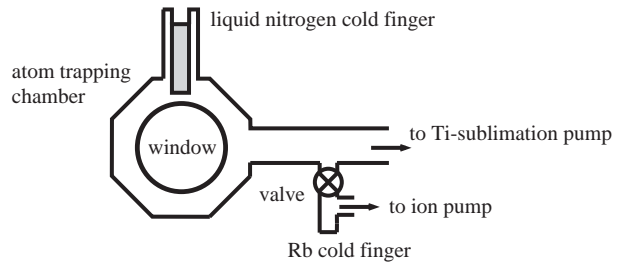,width=1\linewidth,clip=}
\end{center}
\caption[Illustration of Rb apparatus for LIAD]{An illustration of
our apparatus. The MOT was formed in the atom trapping chamber,
and the WLS light entered into the chamber through a
window.}\label{fig:liad1}
\end{figure}

Our stainless steel vacuum chamber consisted of a vapor cell atom
trapping chamber with indium-sealed windows, a liquid
nitrogen-filled cold finger which protruded into the chamber, and
a Rb cold finger at $0^{\circ}$C, as shown in
Fig.~\ref{fig:liad1}. The vacuum in the chamber was maintained by
a Ti-sublimation pump and an ion pump.  We maintained a Rb vapor
in the chamber by slightly opening a valve between the chamber and
a Rb cold finger. This replenished Rb that was pumped out of the
chamber.

The MOT was constructed using a forced dark SPOT technique
\cite{and94b,ket93}: a 4 mm opaque spot was placed in the center
of the path of the repumping laser light, and was imaged onto the
region in the chamber where the trap was formed. Another laser
beam filled the hole in the repumping beam, and was used to
optically pump trapped atoms into a dark state.  This technique
reduced the trap loss rate due to light-assisted, density-dependent collisions between trapped atoms.  The Rb trapping light
was tuned 13 MHz below the $5S_{1/2}, F=2 \rightarrow 5P_{3/2},
F'=3$ transition, and was provided by six 23 mW/cm$^2$, 1.2 cm
diameter laser beams. The 2.7 mW/cm$^2$ repumping laser beam was
tuned 15 MHz below the $F=1 \rightarrow F'=2$ transition, and the
9 mW/cm$^2$ forced optical pumping light was tuned to the $F=2
\rightarrow F'=2$ transition. The number of atoms in the trap was
measured by detecting light scattered by the trapped atoms. This
was done by turning off the $F=2 \rightarrow F'=2$ light for $\sim
50$ ms and filling the hole in the repumping beam with a separate
bypass repumping beam such that the trapped atoms were made bright
by scattering light from the trapping beams.  A fraction of the
light scattered by the trapped atoms was collected and focused
onto a calibrated photomultiplier tube.  Loading rates ($R$) and
MOT loading time constants ($\tau_{\scrs{MOT}}$) were measured by
detecting the number of atoms at sequential points in time as the
trap filled.

The white light used to enhance trap loading was provided by a
fiber optic illuminator, consisting of a halogen bulb with
variable power and a fiber bundle which pointed the light into the
vapor cell.  The coupling of the light from the bulb into the
fiber gave a maximum intensity onto the inner vapor cell wall of
$\sim 10$ W/cm$^2$.  The WLS was switched on and off
electronically.

\begin{figure}     
\begin{center}
\psfig{figure=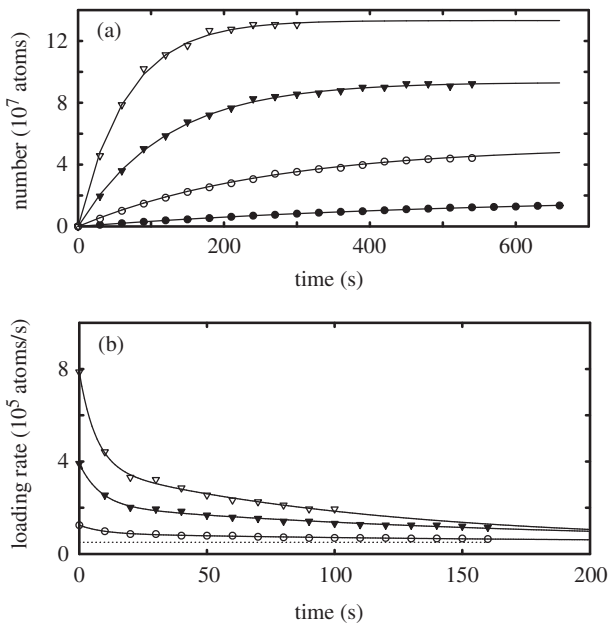,width=1\linewidth,clip=}
\end{center}
\caption[Rb trap loading with and without WLS]{(a) Comparison of trap loading with and
without the additional external white light source. (b) Trend of
the loading rate as a function of time after the WLS is turned
off.  The lower dashed line shows the loading rate before the WLS
was turned on. For both (a) and (b), open triangles represent a
WLS intensity of $\sim10$~W/cm$^2$, closed triangles represent a
WLS intensity of  $\sim5$ W/cm$^2$, and the open circles represent
a WLS intensity of $\sim2$ W/cm$^2$.  The closed circles in (a)
represent loading without the WLS. See Table~\ref{tab:liad1} for
a numerical summary of the data shown in these
plots.}\label{fig:liad2}
\end{figure}

To measure $\tau_{\scrs{MOT}}$, we measured the number of atoms
loaded into the trap as a function of time both with and without
the WLS. The loading curves were exponential in time, as expected
for a trap without light-assisted losses. Typical filling curves
are shown in Fig.~\ref{fig:liad2}(a) for various WLS intensities.
In the figure, the curve representing the fastest filling rate,
with a WLS intensity of $\sim10$ W/cm$^2$, shows a loading time
constant of $\tau_{\scrs{MOT}}=67$ s and a maximum number of $\sim
1.3\times 10^8$ atoms as determined by the exponential fit.
Without the WLS, the loading time constant was
$\tau_{\scrs{MOT}}=538$~s and the maximum number was $\sim2\times
10^7$ atoms. Values of number loaded and loading time constants
for the curves shown in Fig.~\ref{fig:liad2}(a) are given in the
second and third columns of Table~\ref{tab:liad1}\\.

A key factor to consider in optimizing $N\tau$ using the WLS
scheme is the time for the vapor pressure to return to lower
equilibrium values once the WLS is switched off.  We define this
time as the vapor pressure recovery time. A liquid nitrogen cold
finger in the main chamber was used to decrease the Rb vapor
pressure and shorten the recovery time after the WLS was switched
off.  In our cell, the cold finger had little effect on the
background gas pressure, but shortened the recovery time by a
factor of $\sim2$. Furthermore, our experimental timing sequence
consisted of a MOT loading phase with the WLS switched on,
followed by a MOT holding phase, during which the atoms were held
in a MOT with the WLS switched off. This enabled us to keep a
large number of atoms trapped while waiting for the vapor pressure
to decrease before extinguishing the trapping light.

In order to evaluate vapor pressure recovery times, we measured
the dependence of loading rates on
time just after the WLS was switched off. For the data shown in
Fig.~\ref{fig:liad2}(b), the WLS was left on until the Rb partial
pressure reached a saturated level.  The WLS light was then turned
off, and the number of atoms loaded into a MOT in 5 seconds was
repeatedly measured. After each measurement, the MOT light was
kept off for 5 s, and then the MOT started filling again for the
subsequent 5 s filling rate measurement. This set of measurements
indicated the speed at which $\tau_{\scrs{MOT}}$ and the Rb vapor
pressure could recover after the WLS was turned off, and
demonstrated that the recovery time was roughly independent of the
WLS intensity.  The fastest loading rate shown with the WLS on was
$\sim8\times 10^5$ atoms/s, and with the light off was
$\sim2.7\times 10^4$ atoms/s. Each loading rate {\it vs}.\ time
curve in Fig.~\ref{fig:liad2}(b) was fit with a sum of two
decaying exponential curves. The time constants for the loading
rate to return to lower equilibrium values were $\sim8$ s for the
fast recovery time ($\tau_{rec,1}$), and between 113 s and 167 s
for the slower recovery time ($\tau_{rec,2}$).
Table~\ref{tab:liad1} contains a list of recovery times.

\begin{table}[b]     
\begin{center}
\psfig{figure=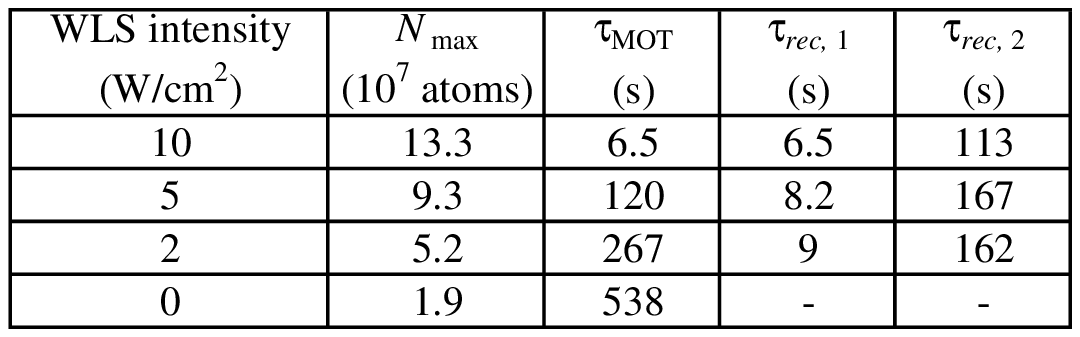,width=1\linewidth,clip=}
\end{center}
\caption[Rb loading and recovery characteristics with induced
desorbtion]{Loading and recovery characteristics of the MOT for
various WLS strengths. $N_{\scrs{max}}$ indicates the maximum
number of atoms that can be loaded into the trap for the
corresponding Rb partial pressure. The MOT loading time constant
is given by $\tau_{\scrs{MOT}}$ and the fast recovery time
constant is listed as $\tau_{rec,1}$. The time $\tau_{rec,2}$ is
the longer time constant in the exponential fits to the data shown
in Fig.~\ref{fig:liad2}(b).}\label{tab:liad1}
\end{table}

To help evaluate the vapor cell performance, the values for
$N_{\scrs{lim}}$ and $\tau_{\scrs{b}}$ were estimated by measuring
$N$ and $\tau_{\scrs{MOT}}$ for various Rb partial pressures.
Experimentally, we varied the Rb partial pressure by adjusting the
intensity of the WLS. We estimated $\tau_{\scrs{b}}\sim700$ s and
$N_{\scrs{lim}}=1.9\times 10^8$ atoms for our operating parameters
by linear extrapolation with our data.

\section{Model}

We now describe a detailed model for determining the numbers and
lifetimes of traps loaded with the WLS to demonstrate the
possibility of increasing $N\tau$ under realistic experimental
conditions. Specifically, this model includes the effects of
finite vapor pressure recovery times and finite loading times.
Without the use of the WLS, and with long loading times, $N\tau$
in a magnetic trap can obtain a maximum optimal value of
\begin{equation}
 (N\tau)_{\scrs{opt}}\equiv N_{\scrs{lim}}\tau_{\scrs{b}}/4\alpha
 \label{eq:liadntau0}
\end{equation}
with $\tau_{\scrs{Rb}}=\tau_{\scrs{b}}$.  We will compare the
performance of a WLS-loaded MOT to $(N\tau)_{\scrs{opt}}$ to
demonstrate the effectiveness of a WLS-loaded MOT.

For a trap loaded with the WLS, calculating $N\tau$ is more
complicated.  We divide the experimental cycle into three time
periods.  During the first period, the MOT is loaded, and the WLS
remains on for the duration of this period. We call this period
the \emph{MOT loading phase}, which has a duration of time $t_1$.
The cycle then enters the \emph{MOT holding phase}, which has a
duration of time $t_2$. In the holding phase, the WLS remains off,
allowing the vapor pressure to recover while continuing to hold a
large fraction of the trapped atoms in the MOT. In the third
period of the cycle, the MOT beams are also turned off and the
atoms are trapped in a magnetic trap. This period begins at time
$t_{\scrs{T}}=t_1+t_2$.

\begin{figure}     
\begin{center}
\psfig{figure=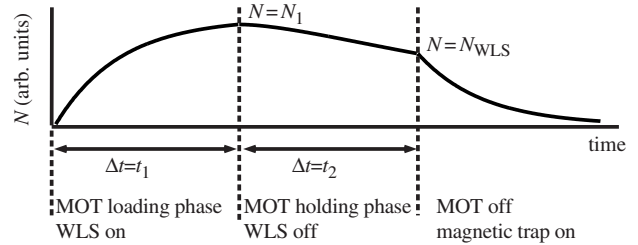,width=1\linewidth,clip=}
\end{center}
\caption[Timing sequence for Rb LIAD experiments and
calculations]{The timing sequence used in the experiment and in
the calculations. The plot is a representation of the number of
trapped atoms as a function of time. The dashed lines separate the
different stages of trap loading and holding. The number of
trapped atoms at the end of the MOT loading and holding phases is
given above the $N$ \emph{vs}.\ time curve.  The states of the MOT
and the WLS are also listed for the time intervals $t_1$ and
$t_2$. After the MOT holding phase, at time
$t_{\scrs{T}}=t_1+t_2$, $N_{\mbox{\scriptsize{WLS}}}$ atoms are
assumed to be loaded from the MOT into a magnetic
trap.}\label{fig:liad3}
\end{figure}

Variables for the number of atoms in the trap can be defined at
the boundaries of the time periods.  At the beginning of the
loading phase, $N=0$. By the end of the loading phase at time
$t_1$, $N_1$ atoms are in the MOT.  The cycle then enters the
holding phase, during which some atoms are lost from the trap due
to collisions with other trapped atoms at a rate that is faster than
the decreasing loading rate into the trap.
We define $N_{\scrs{WLS}}$
to be the number of trapped atoms remaining at the end of this
period. The ``WLS'' subscript emphasizes that this number was
obtained using the WLS.  The cycle then enters the magnetic trap
phase, and $N_{\scrs{WLS}}$ atoms are loaded into the magnetic
trap. Because of the continually decreasing vapor pressure (from
having used the WLS and then turning it off), the number of atoms in the
magnetic trap decays faster than exponentially.  Since we desire to
maximize the number-lifetime product for the magnetic trap, we
define an
\emph{effective lifetime} $\tau_{\scrs{WLS}}$ as the time at
which the number of atoms in the magnetic trap has reached
$(1/e)N_{\scrs{WLS}}$. The entire cycle as described is
illustrated in Fig.~\ref{fig:liad3}.

Our intent in this analysis is to compare
$N_{\scrs{WLS}}\tau_{\scrs{WLS}}$ with both $N\tau$ for
unmodulated Rb pressures at varying loading times and with
$(N\tau)_{\scrs{opt}}$, as defined in Eq.~\ref{eq:liadntau0}.
First, we calculate the number of atoms $N_1$ in the MOT at $t_1$.
At the beginning of the MOT loading phase, the WLS is turned on,
and the loading time constant associated with the Rb partial
pressure quickly drops to a value of $\tau_1$. We thus obtain
\begin{equation}
N_1=\frac{N_{\scrs{lim}}}{1+\tau_1/\tau_{\scrs{b}}}\left[1-\exp{\left(-t_1\left[1/\tau_{1}+1/\tau_{\scrs{b}}\right]\right)}\right]
 \label{eq:liadnwlsmax}
\end{equation}
from the use of Eqs.~\ref{eq:liadloading} and~\ref{eq:liadN0}.

The trapped atoms then enter the holding phase. The WLS is turned off, and the
number of atoms in the MOT is governed by the rate equation
$dN/dt=R(t)-N/\tau_{2}(t)-N/\tau_{\scrs{b}}$, where $\tau_2$ is
the loading time constant associated with the decaying Rb vapor
pressure.  The time dependence of $R$ and $\tau_2$ is made
explicit, since these values depend upon the decreasing Rb vapor
pressure. The loading rate $R(t)$ and the loss rate
$1/\tau_{2}(t)$ are assumed to decay exponentially (with a time
constant of the vapor pressure recovery time) to the steady-state
values $R(t)\rightarrow 0$ and $1/\tau_{2}
\rightarrow 0$ (negligible Rb vapor pressures) as the
vapor pressure recovers. The rate equation is numerically
integrated to determine the number of atoms, $N_{\scrs{WLS}}$,
left in the MOT at time $t_{\scrs{T}}$, the point at which the MOT
is turned off.

We finally must determine the effective lifetime $\tau_{\scrs{WLS}}$
of the magnetic trapping phase of the cycle by
numerically solving the rate equation
$dN/dt=-\alpha(N/\tau_{3}(t)-N/\tau_{\scrs{b}})$.  Here,
$1/\tau_{3}(t)$ has an initial value of $1/\tau_{2}(t_2)$ and
decays exponentially in time to 0 as the vapor pressure continues
to recover. Finally, we can write the number-lifetime product of
the WLS-loaded MOT, designated by $(N\tau)_{\scrs{WLS}}$, as
$(N\tau)_{\scrs{WLS}}=N_{\scrs{WLS}}\tau_{\scrs{WLS}}$.

\section{Results}

\begin{figure}     

\begin{center}
\psfig{figure=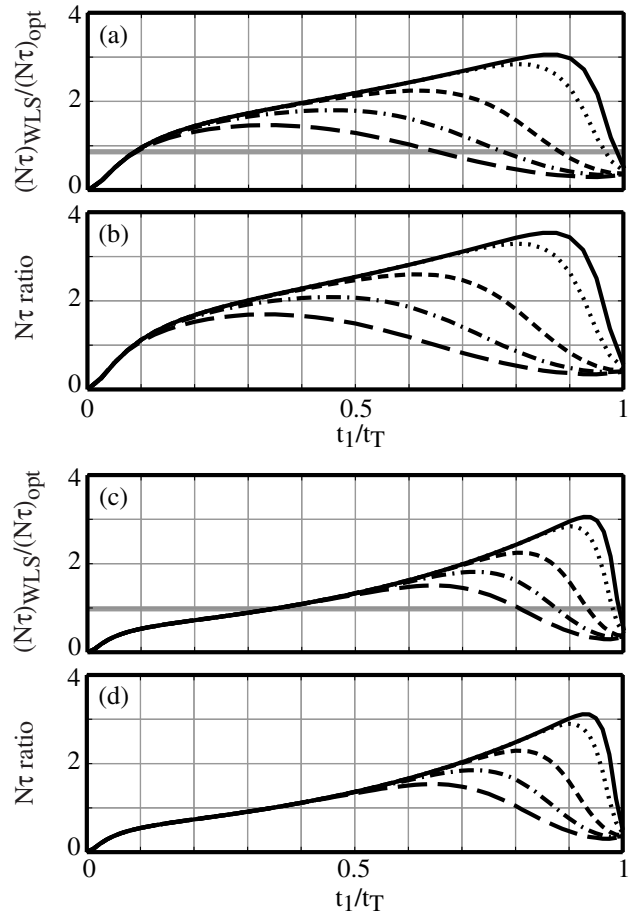,width=1\linewidth,clip=}
\end{center}
\caption[Calculated number-lifetime products with modulated Rb partial pressure]
{ (a) The calculated products $(N\tau)_{\scrs{WLS}}$ in units of
$(N\tau)_{\scrs{opt}} \equiv
N_{\scrs{lim}}\tau_{\scrs{b}}/4\alpha$, the maximum value that can
be achieved with unmodulated vapor pressures, after a total MOT
trapping time of $t_{\scrs{T}}=\tau_{\scrs{b}}$. The Rb-limited
lifetime is assumed to quickly decrease to
$\tau_{1}=\tau_{\scrs{b}}/10$ when the WLS is turned on. The
curves show $(N\tau)_{\scrs{WLS}}$ at varying times $t_{1}$ (as a
fraction of $t_{\scrs{T}}$), the point in the MOT loading cycle at
which the WLS is turned off, with the remaining time in the cycle
($t_{\scrs{T}}-t_{1}$) permitting recovery of the Rb partial
pressure with a time constant of $0.035
\tau_{\scrs{b}}$ (black line), $0.05 \tau_{\scrs{b}}$ (dotted),
$0.1 \tau_{\scrs{b}}$ (short dashed), $0.15
\tau_{\scrs{b}}$ (dashed-dotted), and $0.2 \tau_{\scrs{b}}$ (long dashed).  $N\tau$
for unmodulated vapor pressure is shown as a solid gray line,
after a loading time of $\tau_{T}=\tau_{\scrs{b}}$.  (Note that for this value to equal $(N\tau)_{\scrs{opt}}$,
an infinite loading time would be needed.) (b) The
ratios of the upper curves in (a) to the value of $N\tau$ for
unmodulated vapor pressure after a MOT trapping time of
$\tau_{T}=\tau_{\scrs{b}}$. (c,d) Same as for (a) and (b), with a
total loading time of $t_{\scrs{T}}=2\tau_{\scrs{b}}$. Graphs (c) and (d)
look nearly identical because $N\tau$ for unmodulated
partial pressures is nearly equal to $(N\tau)_{\scrs{opt}}$ after
loading for $t_{\scrs{T}}=2\tau_{\scrs{b}}$.  The four graphs
(a)-(d) shown in this figure
demonstrate that for a given recovery time and MOT trapping time
($t_{\scrs{T}}$), there is an optimal WLS duration ($t_{1}$) the
maximum point on the plotted curves.}
\label{fig:liad4}
\end{figure}

We numerically investigated the performance of the MOT loaded with
the WLS by comparing $(N\tau)_{\scrs{WLS}}$ with $N\tau$ for
unmodulated pressures (Fig.~\ref{fig:liad4}).
Figure~\ref{fig:liad4}(a) shows the number-lifetime product due to
trapping atoms in a MOT for a time $t_{\scrs{T}} =\tau_{\scrs{b}}$
as a function of the fraction of the loading cycle that the WLS is
used. We assume that $N_{\scrs{max}}=N_{\scrs{lim}}/2$ for
unmodulated partial pressures, and an arbitrarily chosen value of
$N_{\scrs{max}} =
N_{\scrs{lim}}/(1+\tau_1/\tau_{\scrs{b}})=(0.9)N_{\scrs{lim}}$
(see Eq.~\ref{eq:liadnwlsmax}), or equivalently
$\tau_{1}=(0.1)\tau_{\scrs{b}}$, for the modulated partial
pressures. Here, the chosen value of $N_{\scrs{max}}$ can not be
set to $N_{\scrs{lim}}$ due to limitations in the numerical
calculations. Fig.~\ref{fig:liad4}(b) shows the same conditions as
Fig.~\ref{fig:liad4}(a), but here we have plotted the ratio of
$(N\tau)_{\scrs{WLS}}$ to $N\tau$ with unmodulated pressures after
a total MOT trapping time of $t_{\scrs{T}}$.

\begin{figure}     

\begin{center}
\psfig{figure=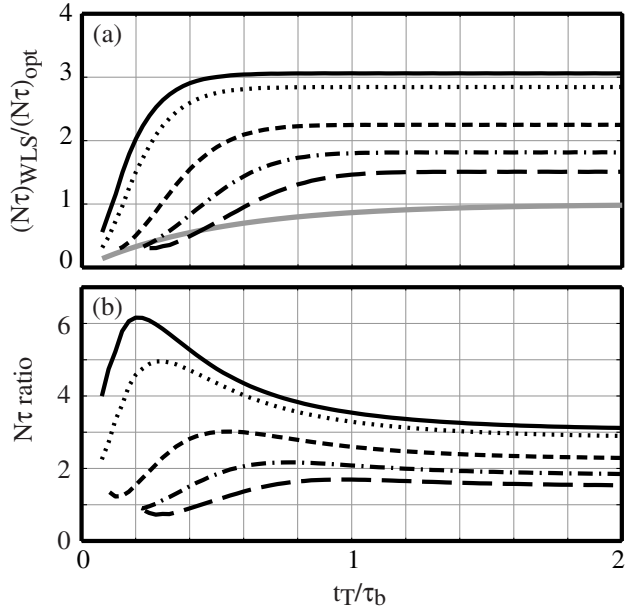,width=1\linewidth,clip=}
\end{center}
\caption[Variation of number-lifetime products
in time, with modulated Rb partial pressure]{(a)  Values of
$(N\tau)_{\scrs{WLS}}$ for WLS loading as a function of
$t_{\scrs{T}}$ (as a fraction of $\tau_{\scrs{b}}$). The limiting
number of atoms that can possibly be loaded into the WLS trap is
assumed to be $(0.9)N_{\scrs{lim}}$ (top curves), or equivalently
$\tau_{1}=\tau_{\scrs{b}}/10$ when the WLS is turned on. The
curves represent recovery times $0.035\tau_{\scrs{b}}$ (black
line), $0.05
\tau_{\scrs{b}}$ (dotted), $0.1 \tau_{\scrs{b}}$ (short dashed), $0.15\tau_{\scrs{b}}$
(dashed-dotted), and $0.2 \tau_{\scrs{b}}$ (long dashed).  The
variation of non-WLS number-lifetime product with time (solid gray
line) is shown.  The calculations assume that for any particular
value of $t_{\scrs{T}}$, the WLS is turned off at the optimal
$t_1$ (see Fig.~\ref{fig:liad4}). The vertical axis is scaled to
$(N\tau)_{\scrs{opt}}$.  (b)  The ratios of the upper curves in
(a) to the value of $N\tau$ for unmodulated vapor pressure at any
given total MOT trapping time $t_{\scrs{T}}$.}
  \label{fig:liad5}
\end{figure}

As suggested by Fig.~\ref{fig:liad4}, the optimum time to leave on the WLS is determined by the maximum
point on a given curve.  In the calculations, the gain in $N\tau$
after using the WLS is less than the maximum possible value of 4
due to the need to allow the vapor pressure to recover before
loading the atoms into a magnetic trap.  The highest values that can be
achieved for $N\tau$ with and without the WLS are plotted against
total loading time $t_{\scrs{T}}$ in Fig.~\ref{fig:liad5} for the
same conditions as in Fig.~\ref{fig:liad4}.  The gain in using the
WLS is again less than the ideal maximum factor of 4 for long
loading times.  However, for short loading times, $N\tau$ for
unmodulated pressures is much lower than $(N\tau)_{\scrs{opt}}$ as
shown by the gray curve in Fig.~\ref{fig:liad5}(a).
Modulated vapor pressures can give substantial benefits in this
regime, as shown by the larger $N\tau$ ratios in
Fig.~\ref{fig:liad5}(b).

As a concrete example of reading the plots given here, we assume
that we have a system that has a vapor pressure recovery time of
$0.035\tau_{\scrs{b}}$.  Thus we are interested in the uppermost curves
in Figs.~\ref{fig:liad4}(a-d) and \ref{fig:liad5}(a,b).  If we
load the vapor cell MOT without modulating the Rb partial
pressure, we can achieve a value of $N\tau \sim 1$ (in units of
$(N\tau)_{\scrs{opt}} \equiv
N_{\scrs{lim}}\tau_{\scrs{b}}/4\alpha$) after loading the trap for
a total time of $t_{\scrs{T}}=2\tau_{\scrs{b}}$, as shown in the
lower (gray) curve of Fig.~\ref{fig:liad5}(a).  However, if we modulate
the Rb pressure with the WLS, we can triple the value of $N\tau$
for the same total MOT trapping time.  To determine the proper
time to remove the WLS light, Fig.~\ref{fig:liad4}(c) indicates
beginning the MOT holding phase $0.12\tau_{\scrs{b}}$ before loading
the magnetic trap (thus $t_{\scrs{1}}=1.88\tau_{\scrs{b}}$)
for optimum trap loading.

\pagebreak

Alternatively, we can shorten the loading time $t_{\scrs{T}}$ to $\sim 0.5
\tau_{\scrs{b}}$, as demonstrated in Fig.~\ref{fig:liad5}(a), and maintain
the same gain in $N\tau$. In doing so, we would not only gain a
factor of 3 in $N\tau$, but we would also increase the repetition
rate of the experiment by as much as a factor of 4.  If instead we
load the experiment for a fixed amount of time in either case
(with or without the WLS), we should look at
Fig.~\ref{fig:liad5}(b) to compare the $N\tau$ products. For a
total MOT trapping time of $t_{\scrs{T}}=0.4\tau_{\scrs{b}}$, we
would achieve over a five-fold gain in $N\tau$ by modulating the
vapor pressure with the WLS.

Experimentally, we were able to obtain an $N\tau$ product of
$(N\tau)_{\scrs{WLS}}=5.0\times 10^9$ atoms$\cdot$s using the WLS
technique, achieved with a MOT loading phase of duration $t_1 =
100$ s and a MOT holding phase of duration $t_2$ = 50 s. This
value of $(N\tau)_{\scrs{WLS}}$ is 2.2 times larger than
$(N\tau)_{\scrs{opt}}=2.3\times 10^9$ atoms$\cdot$s, $95\%$ of the
maximum value of $(N\tau)_{\scrs{opt}}$ for optimized Rb partial
pressure, reached with a loading time of $3\tau_{\scrs{b}}=466$~s.
Without the WLS, the Rb partial pressure was optimized when
$\tau_{\mbox{\scriptsize{MOT}}}=\tau_{\mbox{\scriptsize{Rb}}}/2=\tau_{\scrs{b}}/2$
and $N=N_{\mbox{\scriptsize{lim}}}/2$.  This number of $N\tau$ is
inferred from measurements of $N_{\mbox{\scriptsize{lim}}}$ and
$\tau_{\scrs{b}}$ mentioned previously.

Note that in addition to the gain in $N\tau$, the time to reach
the above value of $(N\tau)_{\scrs{WLS}}$ is 3.1 times faster than
the time to reach the above value of $(N\tau)_{\scrs{opt}}$ (without the WLS),
tripling the repetition rate of experiments. The WLS experimental
technique would be even more beneficial by shortening the recovery
time of the vapor pressure. This might be accomplished by keeping
a larger fraction of the inner surface of the vapor cell at
cryogenic temperatures or through optimization of the surface
adsorption chemistry.

\section{Conclusions and summary}

The use of LIAD to enhance loading of vapor cell MOTs may be
applicable to other atomic species.  Lithium vapor cells, for
instance, are difficult to work with due to the high temperatures
needed to create a substantial Li vapor.  Yet if LIAD were to work
well with Li adsorbed on stainless steel, or between co-adsorbed
Li atoms on a surface, a Li vapor cell MOT would be practical.
Although the effect has not yet been quantitatively explored as it
has been for Rb, we observed a LIAD induced increase in the
loading rate into a Cs MOT in a Cs vapor cell with aluminum walls.
In general, when first using the LIAD technique,
the WLS intensity should be raised incrementally to monitor the
loading time constant. When the WLS loading time constant
becomes too short ($< 10$ s) the vapor pressure can
potentially become high enough that atoms may re-adsorb onto cold
chamber windows and may possibly form small clusters of atoms.

In summary, we have demonstrated that the technique of non-thermal
light induced atom desorption can be used to effectively increase
the number of atoms that can be loaded into a vapor cell MOT. This
technique benefits atom trapping experiments where large numbers
of atoms and long trap lifetimes are crucial.



\begin{thebibliography}{99}

\bibitem[*]{BPApresentaddress}  Current address: JILA, Campus Box
440, University of Colorado, Boulder, CO, 80309-0440.

\bibitem{and95}   M.H.~Anderson, J.R.~Ensher, M.R.~Matthews, C.E.~Wieman,
E.A.~Cornell, {\it Science} {\bf 269}, 198 (1995).

\bibitem{dav95} K.B.~Davis, M.-O.~Mewes, M.R.~Andrews, N.J.~van Druten,
D.S.~Durfee, D.M.~Kurn, W.~Ketterle, {\it Phys.~Rev.~Lett.} {\bf 75},
3969 (1995).

\bibitem{bra95}  C.C.~Bradley, C.A.~Sackett, J.J.~Tollett, R.G.~Hulet,
{\it Phys.~Rev.~Lett.} {\bf 75}, 1687 (1995); C.C.~Bradley,
C.A.~Sackett, R.G.~Hulet, {\it Phys.~Rev.~Lett.} {\bf 78}, 985 (1997).

\bibitem{fri98}  D.G.~Fried, T.C.~Killian, L.~Willmann,
D.~Landhuis, S.C.~Moss, D.~Kleppner, T.J.~Greytak, {\it
Phys.~Rev.~Lett.} {\bf 81}, 3811 (1998).


\bibitem{mon90} C.~Monroe, W.~Swann, H.~Robinson, C.~Wieman, {\it
Phys.~Rev.~Lett.} {\bf 65}, 1571 (1990).

\bibitem{raa87JOSAB89}  E.L.~Raab, M.~Prentiss, A.~Cable, S.~Chu,
D.E.~Pritchard, {\it Phys.~Rev.~Lett.} {\bf 59}, 2631 (1987); also
see {\it J.~Opt.~Soc.~Am.~B} {\bf 6}, No.~11 (1989).

\bibitem{mya96} C.J.~Myatt, N.R.~Newbury, R.W.~Ghrist, S.~Loutzenheiser,
C.E.~Wieman, {\it Opt.~Lett.} {\bf 21}, 290 (1996).

\bibitem{and99} B.P.~Anderson and M.A.~Kasevich, {\it
Phys.~Rev.~A} {\bf 59}, R938 (1999).

\bibitem{bon90} A.M.~Bonch-Bruevich, T.A.~Vartanyan, Yu.M.~Maksimov,
S.G.~Przhibel'ski\u{i}, V.V.~Khromov, {\it Sov.~Phys.~JETP} {\bf 70},
993 (1990).

\bibitem{meu94xu87mar94} M.~Meucci, E.~Mariotti, P.~Bicchi,
C.~Marinelli, L.~Moi, {\it Europhys.~Lett.} {\bf 25}, 639
(1994).~See also J.~Xu, M.~Allegrini, S.~Gozzini, E.~Mariotti,
L.~Moi, {\it Opt.~Comm.} {\bf 63}, 43 (1987); and E.~Mariotti,
S.~Atutov, M.~Meucci, P.~Bicchi, C.~Marinelli, L.~Moi, {\it
Chem.~Phys.} {\bf 187}, 111 (1994).

\bibitem{adsorbgeneral} For a general discussion of adsorption and
desorption, see Morrison, S.~Roy, {\it The Chemical Physics of
Surfaces}, second edition (Plenum Press) 1990; and R.~Masel, {\it
Principles of Adsorption and Reaction on Solid Surfaces} (John
Wiley \& Sons, Inc.) 1996.

\bibitem{wal94} T.G.~Walker and P.~Feng, {\it Advances in Atomic,
Molecular, and Optical Physics} {\bf 34}, 125 (1994).

\bibitem{and94b} M.H.~Anderson, W.~Petrich, J.R.~Ensher, E.A.~Cornell,
{\it Phys.~Rev.~A} {\bf 50}, R3597 (1994).

\bibitem{ket93} W.~Ketterle, K.B.~Davis, M.A.~Joffe, A.~Martin,
D.E.~Pritchard, {\it Phys.~Rev.~Lett.} {\bf 70}, 2253 (1993).


\end{thebibliography}
\end{document}